\documentclass[aip,letter,twocolumn,amsmath,amssymb]{revtex4}
\usepackage{graphicx}
\begin{document}
\title{Low Ghz loss in sputtered epitaxial Fe}
\author{C. Scheck, L. Cheng, W.E. Bailey}
\affiliation{Materials Science and Engineering Program, Dept. of
Applied Physics and Applied Mathematics, Columbia University, 500 W
120th St, New York, NY 10027}
\homepage{http://magnet.ap.columbia.edu}

\date{\today}

\begin{abstract}
We show that sputtered, pure epitaxial iron films can have
high-frequency loss as low as, or lower than, any known metallic
ferromagnetic heterostructure.  Minimum 34 Ghz ferromagnetic
resonance (FMR) linewidths of 41 $\pm$ 2 Oe are demonstrated, some
$\sim$ 5-10 \% lower than the previous minimum reported for
molecular beam epitaxially (MBE) deposited Fe. Intrinsic and
extrinsic damping have been separated over 0-40 Ghz, giving a lower
bound for intrinsic LL(G) relaxation rates of $\lambda$ or
$G=\textrm{85}\pm\textrm{5 MHz}$ ($\alpha$ = 0.0027 $\pm$ 0.0001)
and extrinsic $\eta\sim\textrm{50 Mhz}$.  Swept frequency
measurements indicate the potential for integrated frequency domain
devices with $Q>100$ at 30-40 Ghz.
\end{abstract}

\maketitle
\section{Introduction}
Low damping $\alpha$, or relaxation rates $\lambda$, are essential
for high frequency applications of magnetic heterostructures.
Nanoscale spin electronic sensors operating above 1 Ghz have
signal-to-noise ratios (SNR) which depend inversely on the damping
constant $\alpha$ and are independent of spin transport
parameters.\cite{neilsmith-snr} Integrated magnetic frequency domain
devices have frequency linewidths ($\Delta\omega$/2$\pi$) limited
fundamentally by the Landau-Lifshitz-(Gilbert) relaxation rate
$\lambda(=G)=\alpha\gamma\:M_{s}$,\cite{patton13} where $\gamma$ is
the gyromagnetic ratio.  It is timely to determine how low
relaxation rates can be made in a ferromagnetic thin film,
particularly using widely accessible deposition techniques such as
sputtering.

Relaxation processes expressed phenomenologically in
$\alpha$\cite{pimm} can be divided into extrinsic and intrinsic
types. Extrinsic damping results from microstructure; intrinsic
damping results from spin-orbit
coupling.\cite{kambersky-microscopic} The two effects can be
separated through variable-frequency ferromagnetic resonance
measurements (FMR), through $\Delta H_{pp}=\Delta
H_{0}+(2/\sqrt{3})\alpha/\gamma$.\cite{heinrich-review} $\alpha$ in
this context expresses intrinsic processes, and $\Delta H_{pp}$
expresses inhomogeneous broadening due to e.g. line
defects.\cite{heinrich-woltersdorf}

The lowest overall linewidths have been seen in the ultrathin
molecular beam epitaxially (MBE) deposited Fe films of Prinz, with a
35 Ghz $\Delta H_{pp}=\textrm{45 Oe}$ (1.29 Oe/Ghz) seen in
ultrathin Fe(100) deposited on ZnSe(100) epilayers.\cite{prinz-fe}
Intrinsic and extrinsic losses were not separated in the prior work,
carried out at a single frequency.  Fe also possesses the lowest
known {\it intrinsic} damping constant of any metallic ferromagnet,
with a range of $\lambda$ quoted as $\lambda=\textrm{70-140 Mhz}$
($\alpha=\textrm{0.002-0.004})$ in FMR measurement to 40
Ghz.\cite{l-b-lambda} Variable frequency FMR estimates of $\alpha$
over this range, through $\partial \Delta H/\partial\Delta\omega$,
have typically uncovered values of $\textrm{100 Mhz}\leq\lambda\leq$
140 Mhz in high-quality MBE\cite{heinrich-woltersdorf,farle-fmr} or
sputtered films.\cite{lubitz-jap-cu-fe}

In this work, we report UHV sputtered epitaxial pure
Fe(100)(15nm)/Ti(2nm) films on MgO(100) which show FMR linewidths of
$\Delta H_{pp}$ = 41 $\pm$ 2 Oe at 34 Ghz (1.20 Oe/Ghz), some 5-10\%
lower than the narrowest linewidths seen to date in MBE deposited
films.  Variable frequency 0-40 Ghz FMR indentifies $\lambda$ = 85
$\pm$ 5 Mhz ($\alpha=\textrm{0.0027}\pm\textrm{0.0001}$) and $\Delta
H_{0}\sim\textrm{6}\pm\textrm{2 Oe}$ for these thin films; a role of
eddy current damping is identified in $\alpha$ of thicker Fe films.
Swept-frequency measurements demonstrate the potential for
field-tunable $\sim$ 35 Ghz filters with $Q>100$, an order of
magnitude better than achieved previously in Fe.

\section{Experimental}

Fe (8-75 nm) thin films were deposited on polished MgO(001)
substrates using dc magnetron UHV sputtering at a base pressure of
3.0$\times$10$^{-9}$ Torr.  Pressures immediately prior to
deposition after sample introduction were typically
1.0$\times$10$^{-8}$ Torr.  Substrates were held at 200 $^{\circ}$C
during sputter deposition, at 4$\times$10$^{-3}$ Torr {\it in-situ}
getter-purified Ar, 300 W power for 2 inch targets, and 10 cm
target-substrate spacing.  Growth rates of $\sim$6 {\AA}/s were
measured by a quartz crystal microbalance and {\it ex-situ}
profilometry.  Films were capped with 2 nm sputtered Ti to protect
the surface from oxidation.  Rocking curve half widths measured for
50 nm films were very low, only 0.5 $^{\circ}$, and roughly
independent of deposition temperature over the range 200-300
$^{\circ}$C.  Results for ion beam sputtered Ni$_{81}$Fe$_{19}$(48
nm) are plotted for comparison; see Ref. \cite{reidy-apl} for
deposition conditions.

Broadband FMR measurements were carried out using microwave
frequencies in the range 4-40 Ghz generated by a synthesized sweep
generator operating in cw mode.  Microwaves were applied to the
samples through a coplanar waveguide (CPW) for the range 4-18
Ghz\cite{pimm} and a shorted K-band rectangular waveguide for higher
frequencies, with diode detector in transmission and reflection,
respectively.  Derivative spectra $\Delta \chi^{''}/\Delta H$ were
recorded using ac field modulation ($<\pm$2 G) and lock-in
detection.\cite{heinrich-advphys} Swept-field and swept-frequency
measurements were both carried out, at room temperature.

\section{Results}
\begin{figure}[htb]
\includegraphics[width=\columnwidth]{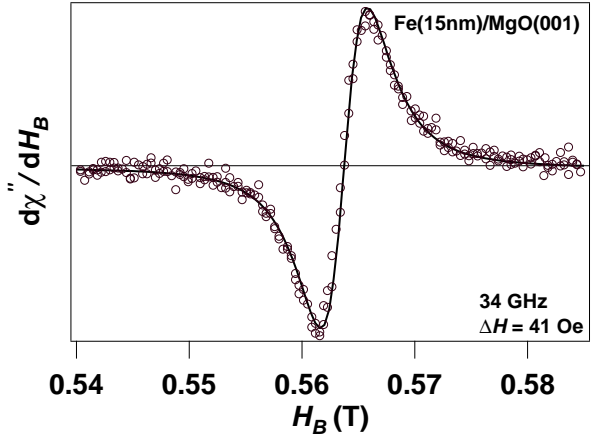}
\caption{34 Ghz FMR spectrum, with Lorenzian fit, for epitaxially
sputtered MgO(001)/Fe(15 nm)/Ti(2 nm).  See text for
details.}\label{FMR}
\end{figure}

A representative FMR spectrum for thin (8 or 15 nm) Fe films at 34
Ghz is shown in Fig. \ref{FMR}.  The film is measured with $H$
applied along the $<$110$>$ hard axis, along MgO$<$100$>$. The
derivative spectrum is shown to be symmetric, with Lorenzian fit
indicated, and peak-to-peak linewidth measured of $\Delta
H_{pp}=\textrm{41}\pm\textrm{2 Oe}$.

\begin{figure}[htb]
\includegraphics[width=\columnwidth]{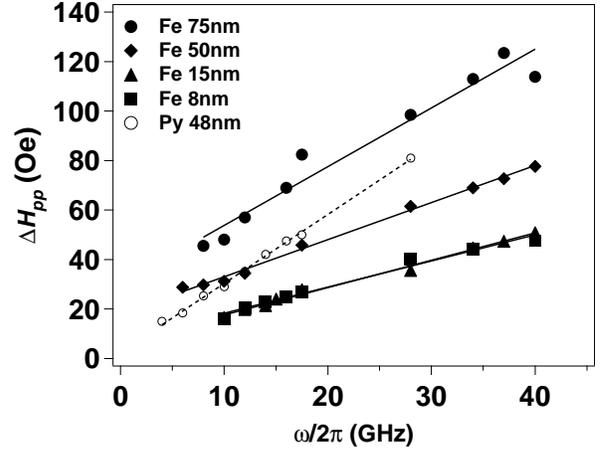}
\caption{Frequency dependent peak-to-peak FMR linewidths $\Delta
H_{pp}$ for epitaxially sputtered MgO(100)/Fe(t), $\textrm{8
nm}<t<\textrm{75 nm}$, with linear fits to extract $\alpha$.  Data
from polycrystalline SiO$_{2}$/Ni$_{81}$Fe$_{19}$(48 nm) are plotted
for comparison. }\label{DH}
\end{figure}

Peak-to-peak FMR linewidths $\Delta H_{pp}$ versus frequency
$\omega/2\pi$ were plotted for all samples (Fig. \ref{DH}) to
determine the Landau-Lifshitz-Gilbert (LLG) damping constant
$\alpha$ and the inhomogeneous broadening $\Delta H_{0}$.  From the
slope $\partial \Delta H/\partial\Delta\omega$, we find a minimum
$\alpha=\textrm{0.0027}\pm\textrm{0.0001}$ for thin Fe ($<$15 nm).
$\alpha=\textrm{0.0075}$ is measured for Ni$_{81}$Fe$_{19}$(48nm),
consistent with the lower bound of typical values and characteristic
of high quality films.  Relaxation rates $\lambda$ are converted
from $\alpha$ measurements using
$g_{eff}=\textrm{2.09}$\cite{l-b-lambda} and $4\pi
M_{s}^{Ni_{81}Fe_{19}}=\textrm{10.6 kG}$, 4$\pi
M_{s}^{Fe}=\textrm{21.6 kG}$, and plotted for comparison.  $\lambda$
reaches a minimum of 85 $\pm$ 5 Mhz for epitaxial Fe and 120 $\pm$
10 Mhz for Ni$_{81}$Fe$_{19}$.  Inhomogeneous broadening is
negligible for Ni$_{81}$Fe$_{19}$, with $\Delta
H_{0}=\textrm{2}\pm\textrm{2 Oe}$, and reaches a minimum of $\Delta
H_{0}=\textrm{6}\pm\textrm{2 Oe}$ for 15 nm Fe.

\begin{figure}[htb]
\includegraphics[width=\columnwidth]{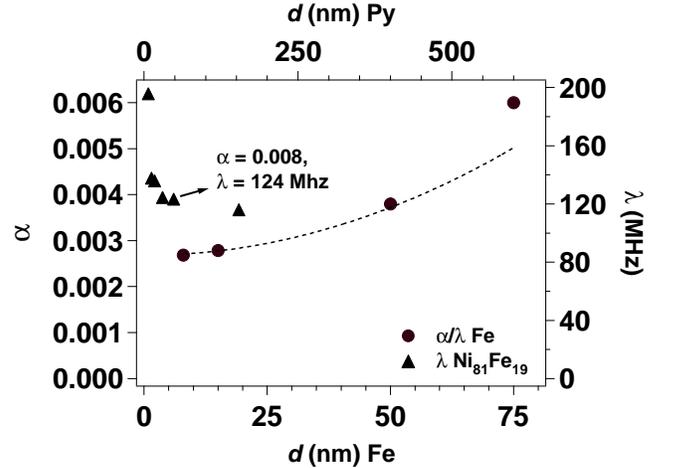}
\caption{Extracted damping constant $\alpha$ (left) and relaxation
rate $\lambda$ for epitaxial Fe films.  The dashed line shows a
calculated contribution of eddy currents to $\lambda$ (Eq.
\ref{eddy}).  $\lambda$ for Ni$_{81}$Fe$_{19}$ is plotted for
comparison.  See text for details.}\label{damping}
\end{figure}

An increasing trend in $\alpha$ with thickness can be seen for Fe
films thicker than 15 nm.  We have compared the increase in
$\lambda$ with a standard theory of eddy current damping,\cite{lock}
which predicts a quadratic increase in Gilbert-type (proportional to
$\omega$) linewidth with film thickness $t$:

\begin{equation}\label{eddy}
\lambda_{\textrm{eddy}}={\sigma\over 12}(4\pi\gamma
M_{s})^2{\left(t\over c\right)}^2
\end{equation}

where $\sigma$ is the conductivity ($hz$ for cgs units) and $c$ the
speed of light in vacuum. $\sigma$ values used for Fe and
Ni$_{81}$Fe$_{19}$ were respectively 9.11$\times$10$^{16}$ hz
($\rho$ = 10 $\mu\Omega$.cm from four-point-probe measurement) and
4.5$\times$10$^{16}$ hz ($\rho$ = 20 $\mu\Omega$.cm). Order of
magnitude agreement is found with the increase in $\lambda$ for Fe
films to 75 nm; the Ni$_{81}$Fe$_{19}$ data are plotted with
thickness scale compressed by the ratio of the prefactors for the
two materials ($\sim 8$), indicating an expected delayed onset of
eddy-current damping ($>$ 200 nm) for Ni$_{81}$Fe$_{19}$.  A
thickness-dependent increase of the inhomogeneous term $\Delta
H_{0}$ from 6 to 30 Oe with increasing Fe thickness may originate in
a higher concentration of strain-relaxing dislocations for thicker
films.\cite{heinrich-woltersdorf}

The advantage of swept-$\omega$ FMR measurement for extracting total
relaxation rates $\eta$ has been pointed out by
Patton.\cite{patton13} $\eta$ can be measured independent of
geometry as

\begin{equation}\label{wpp}
\Delta\omega_{pp}=(2/\surd3)\eta,
\end{equation}
\cite{patton13}where $\eta=1/\tau$, giving the decay time as
$\exp{-\eta t}$ in a time-domain experiment,\cite{pimm} and where
$\eta_{G}=2\pi\lambda$ in the absence of extrinsic relaxation.
$\lambda$ can be estimated in the intrinsic limit as
$\lambda=(\sqrt{3}/2)\gamma \Delta H_{pp}$.  We approximate
extrinsic and intrinsic relaxation rates as
$\eta\simeq\eta_{0}+\eta_{G}$ with $\eta_{0}=\surd3\gamma\Delta
H_{0}$ measured in the absence of $\alpha$.

\begin{figure}[htb]
\includegraphics[width=\columnwidth]{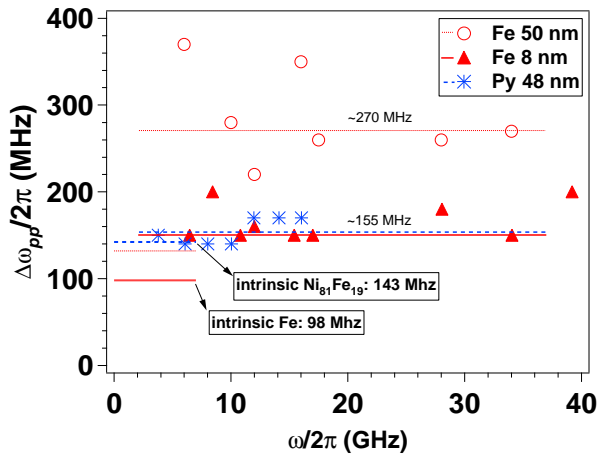}
\caption{Swept-frequency FMR linewidths $\Delta \omega_{pp}/2\pi$
for 8 nm and 50 nm epitaxial Fe; 48 nm (asterix) Ni$_{81}$Fe$_{19}$
is shown for comparison.  Values of relaxation rate $\Delta
\omega_{pp}/2\pi$ from Gilbert damping only are
indicated.}\label{Df}
\end{figure}

Figure \ref{Df} shows a plot of the peak-to-peak swept-$f$ FMR
linewidths versus frequency for Fe (8 and 50 nm) and
Ni$_{81}$Fe$_{19}$ (48 nm) films.  It can be seen that the
Ni$_{81}$Fe$_{19}$ films follow the intrinsic limit quite well, with
$\Delta \omega_{pp}/2\pi$ = 129$\cdot$(2/$\surd$3) = 149 Mhz
(theoretical) approximated to 16 Ghz. However, the inhomogeneous
term is appreciable in 8 nm Fe; $\Delta H_{0}=6\pm\textrm{2 Oe}$
translates to $\eta_{0}$ = 191 $\pm$ 60 Mhz and
$\Delta\omega_{pp}^{0}$/2$\pi$ = (191 Mhz/2$\pi)\cdot($2/$\surd$3) =
35 $\pm$ 10 Mhz, comparable to the observed $\Delta
\omega_{pp}/2\pi-2\lambda/\sqrt{3}\simeq$ 50 $\pm$ 5 Mhz.

\

\section{Discussion}

The observed low extrinsic relaxation rates are a plausible result
of the excellent crystalline quality in the ultrathin epitaxial
sputtered Fe films. Inhomogeneous broadening is more typically
measured on the order of $\Delta H_{0}=\textrm{50
Oe}$,\cite{gilbert-g} compared with the best of $\Delta
H_{0}=\textrm{6 Oe}$ seen here. X-ray diffraction rocking curves of
the (200) peak on our films show full-width-half-maxima (FWHM) as
low as 0.6$^{\circ}$; more standard values for seeded epitaxy in
sputtering for this system are 1.1$^{\circ}$.\cite{parkin-harp}
Moreover, easy-axis ($<$100$>$) coercivities $H_{c}$, measured by
VSM, are 2.1 Oe compared with 3.7 Oe in Ref. \cite{prinz-fe} in
thinner films (50 nm vs. 320 nm for MBE). The inhomogeneous
linewidth is the lowest we are aware of in Fe films.

Finally we comment on applications.  Favorable epitaxial structures
in sputtered Fe/MgO/Fe junctions have resulted in very high
tunneling magnetoresistance;\cite{yuasa} our results indicate that
low $\alpha$ and high $\Delta R/R$ may coexist.  Additionally, the
low 35 Ghz frequency linewidths seen in our epitaxial Fe films could
translate directly to high half-power $Q$ in a frequency domain
device.  One example is a tunable bandstop filter based on FMR. We
see $\omega/\Delta\omega_{1/2}=140$ in our films, roughly an order
of magnitude higher than that realized to date in Fe device
structures.\cite{kuanr-apl}

\section{Acknowledgements}
We thank Z. Frait for helpful discussions.  This work was supported
by the Army Research Office under contracts DA-ARO-W911NF0410168,
DAAD19-02-1-0375, and 43986-MS-YIP, and has used the shared
experimental facilities that are supported primarily by the MRSEC
program of the National Science Foundation under NSF-DMR-0213574.



\end{document}